\documentclass[a4paper,12pt]{spieman}  % use this instead for A4 paper
\usepackage{amsmath,amsfonts,amssymb}
\usepackage{graphicx}
\usepackage{setspace}
\usepackage{tocloft}
\usepackage{bbm}
\usepackage{bm}
\usepackage{nicefrac}
\usepackage{mathtools}
\usepackage{verbatim}
\usepackage{microtype} 
\usepackage{listings}
\usepackage{array}
\usepackage{wrapfig}
\usepackage{textcomp}
\usepackage{url}            % simple URL typesetting
\usepackage{booktabs}       % professional-quality tables

\newcommand{\RR}{\mathbb{R}}

\newcommand{\CC}{\mathbb{C}}

\newcommand{\R}{\mathcal{R}}

\newcommand{\Abf}{\mathbf{A}}

\newcommand{\bbf}{\mathbf{b}}

\newcommand{\xbf}{\mathbf{x}}

\newcommand{\zbf}{\mathbf{z}}

\DeclareMathOperator*{\argmin}{arg\,min}
\newcommand{\defeq}{\vcentcolon=}
\newcommand\norm[1]{\left\lVert#1\right\rVert}

\def\*#1{\mathbf{#1}}
\def\BibTeX{{\rm B\kern-.05em{\sc i\kern-.025em b}\kern-.08em
    T\kern-.1667em\lower.7ex\hbox{E}\kern-.125emX}}

\usepackage{algorithmic,algorithm}
\usepackage{tikz}

\title{Learnable Optimization-Based Algorithms for Low-Dose CT Reconstruction }

\author[a]{Daisy Chen}
\affil[a]{Xinyang Normal University, NanHu Road 237, Xin Yang, China, 464000}

\cftpagenumbersoff{figure}
\cftpagenumbersoff{table} 
\begin{document} 
\maketitle

\begin{abstract}
Low-dose computed tomography (LDCT) aims to minimize the radiation exposure to patients while maintaining diagnostic image quality. However, traditional CT reconstruction algorithms often struggle with the ill-posed nature of the problem, resulting in severe image artifacts. Recent advances in optimization-based deep learning algorithms offer promising solutions to improve LDCT reconstruction. In this paper, we explore learnable optimization algorithms (LOA) for CT reconstruction, which integrate deep learning within variational models to enhance the regularization process. These methods, including LEARN++ \cite{zhang2022learn++} and (MAGIC) \cite{xia2021magic}, leverage dual-domain networks that optimize both image and sinogram data, significantly improving reconstruction quality. We also present proximal gradient descent and ADMM-inspired networks, which are efficient and theoretically grounded approaches. Our results demonstrate that these learnable methods outperform traditional techniques, offering enhanced artifact reduction, better detail preservation, and robust performance in clinical scenarios.

\end{abstract}

% Include a list of up to six keywords after the abstract

% Include email contact information for corresponding author
% {\noindent \footnotesize\textbf{*}Wanyu Bian,  }

\begin{spacing}{2}   % use double spacing for rest of manuscript

\section{Introduction}\label{sec:intro}  

Computed Tomography (CT) is a cornerstone of modern medical diagnostics, providing detailed cross-sectional images of the human body through X-ray measurements taken from multiple angles. It is indispensable for detecting pathological abnormalities such as tumors, vascular diseases, lung nodules, internal injuries, and bone fractures, and plays a crucial role in guiding clinical procedures like interventions, radiation therapies, and surgeries. However, the high doses of X-rays used in conventional CT scans pose significant health risks, including the development of radiation-induced cancers and other genetic disorders, which necessitate minimizing patients' exposure to radiation.

To address this, Low-Dose CT (LDCT) techniques have been developed, aiming to reduce the X-ray dose while preserving diagnostic image quality. One prominent approach within LDCT is Sparse-View CT, which reduces the number of X-ray projections to lower the radiation dose. Despite its benefits, Sparse-View CT presents a significant challenge: the severe undersampling of sinogram data leads to an ill-posed reconstruction problem. Standard algorithms like filtered-back-projection (FBP) applied to Sparse-View CT data result in images with severe artifacts, rendering them unreliable for clinical use.

Variational methods have emerged as a significant class of mathematical approaches for tackling this reconstruction problem. These methods frame image reconstruction as a minimization problem, consisting of a penalty term that measures the discrepancy between the reconstructed image and the observed data and a regularization term that enforces prior knowledge or smoothness in the image. The optimization process seeks the minimizer of this objective function, producing the reconstructed image. However, the regularization functions commonly used, such as Total Variation (TV), often fall short in capturing the fine details of medical images, limiting their effectiveness in real-world applications.

Deep learning (DL) methods for image processing \cite{ding2024confidence,ni2024earnings,bian2024diffusion,ding2024llava,li2024vqa,zhang2020multiscale,zhang2022extra,yukun2019deep} have evolved rapidly, offering significant improvements over traditional techniques. One of the most successful DL-based approaches for CT reconstruction is known as unrolling, which mimics traditional optimization schemes like proximal gradient descent used in variational methods but replaces handcrafted regularization with deep networks \cite{chen2021variational}. These deep networks are utilized to extract features from images or sinograms, enhancing the reconstruction process. Recently, dual-domain methods have emerged, leveraging complementary information from both image and sinogram domains to further improve reconstruction quality. Despite these advancements, DL-based methods face challenges due to their lack of theoretical interpretation and practical robustness. They are often memory inefficient and prone to overfitting, as they superficially mimic optimization schemes without ensuring convergence and stability guarantees. Additionally, convolutional neural networks (CNNs) have been successfully applied to sparse-view and low-dose data, projection domain synthesis, post-processing, and prior learning in iterative methods, demonstrating better performance than analytical approaches.

Recently, a new class of DL-based methods known as learnable optimization algorithms (LOA) has been developed for image reconstruction with mathematical justifications and convergence guarantees, showing promising advancements. LOAs are also largely explored in other medical image reconstruction problems\cite{bian2021optimization}. These methods originate from a variational model where the regularization is parameterized as a deep network with learnable parameters, leading to a potentially nonconvex and nonsmooth objective function. LOAs aim to design an efficient and convergent scheme to minimize this objective function, resulting in a highly structured deep network.  The parameters of this network are inherited from the learnable regularization and are adaptively trained using data while retaining all convergence properties. This approach has been applied to develop dual-domain sparse-view CT reconstruction methods, using learnable regularizations for both image and sinogram domains. These methods unroll parallel subnetworks to extract complementary information from both domains. 

%However, existing works often use convolutional neural networks (CNNs) to model regularizers, which limits the representation power as they only explore local image features. This is inadequate for medical imaging, which demands high-quality images. Moreover, many deep networks for image reconstruction are treated as black-boxes, making them difficult to interpret, and they often lack mathematical justifications and convergence guarantees. LOA have been explored recently to address these issues, boosting reconstruction quality through adaptive non-local feature regularizers and computational efficiency, as safeguards are only computed when necessary. These models retain convergence guarantees while improving reconstruction quality over existing methods.

This paper introduces several different optimization-based algorithms to solve CT reconstruction problem.  Section \ref{sec:intro} introduces the importance and recent developments of CT reconstruction. Section \ref{sec:recon} introduces the variational model for CT reconstruction. Section \ref{sec:opt} introduces a brief overview of the model-based CT reconstruction methods using deep learning.

\section{CT Reconstruction Model}\label{sec:recon}
CT reconstruction is a classic inverse problem that can be formulated variationally and represented as the following optimization problem:
\begin{equation} \label{eq:ls}
    \min_\xbf  \frac{1}{2} \|  \Abf \xbf - \bbf \|^2_2 + \mu \R(\xbf),
\end{equation}

where $\xbf \in \CC^n$ is the CT image to be reconstructed, consisting of $n$ numbers of pixels, and $\bbf \in \CC^m$ denotes the corresponding sinogram measurement data. $\Abf$ is the Radon transform. The data fidelity term $ \frac{1}{2} \|  \Abf \xbf - \bbf \|^2_2$ enforces physical  consistency between the reconstructed image $\xbf$ and the  sinogram measurements $\bbf$. The regularization operator $\R: \CC^n \to \RR$ emphasizes the sparsity of the MRI data or low rankness constraints. The regularizer provides prior information to avoid overfitting the data fidelity term. The weight parameter $\mu > 0$  balances data fidelity term and regularization term.    

\section{Optimization-based network unrolling algorithms for CT reconstruction}\label{sec:opt}

\subsection{Gradient Descent Algorithm Inspired Network}

\subsubsection{LEARN++: Recurrent Dual-Domain Reconstruction Network}
LEARN++: Recurrent Dual-Domain Reconstruction Network \cite{zhang2022learn++} is a dual-domain extension of the state-of-the-art LEARN model \cite{8290981} for compressed sensing (CS) computed tomography (CT). Unlike traditional CS methods that rely on handcrafted prior regularizers and existing iteration unrolling methods that only involve projection data in the data consistency layer, LEARN++ integrates two parallel and interactive subnetworks. These subnetworks perform image restoration and sinogram inpainting simultaneously on both the image and projection domains, fully exploring the latent relationships between projection data and reconstructed images. This model is designed as a cascaded network with successive blocks, each containing the interactive subnetworks. Experimental results show that LEARN++ achieves competitive qualitative and quantitative results, excelling in artifact reduction and detail preservation compared to several state-of-the-art methods.

\subsubsection{Manifold and Graph Integrative Convolutional Network (MAGIC)}
Manifold and Graph Integrative Convolutional Network (MAGIC) \cite{xia2021magic} presents a new architecture combining traditional convolutional neural networks (CNNs) with graph-based methods for improving LDCT image quality. The proposed MAGIC model utilizes spatial convolutions to capture local features and graph convolutions to integrate non-local topological structures from the manifold space, which helps in denoising while preserving critical image details. Key innovations include combining manifold learning with graph-based methods to enhance both image quality and robustness. Extensive experiments demonstrate that MAGIC outperforms state-of-the-art techniques in terms of artifact reduction and detail preservation.

\subsection{Proximal Gradient Descent Algorithm Inspired Networks}

\subsubsection{LDA-LS}
 Learned Descent Algorithm with a Line Search strategy (LDA-LS) \cite{zhang2021nonsmooth} addresses the limitations of the deep neural network architectures based on unrolling optimization algorithms for image reconstruction, which often lack convergence guarantees and interpretability, hindering clinical applications. 
The LDA-LS algorithm addresses \eqref{eq:ls} by replacing $\R$ with a nonsmooth, nonconvex regularization term, which is formulated using a convolutional neural network (CNN) followed by a sparsity-promoting feature selection norm \cite{chen2020learnable}. By learning a nonsmooth, nonconvex regularizer, LDA-LS is able to generate high-quality LDCT images with minimal computational cost.

\subsubsection{Efficient Learned Descent Algorithm (ELDA) }
Efficient Learned Descent Algorithm (ELDA) \cite{zhang2024provably} employs an adaptive non-local feature mapping and associated regularizer to enhance reconstruction quality. The problem that ELDA solves can also be formulated as \eqref{eq:ls}, where the regularization term consists of two parts:
\begin{equation}\label{eq:r}
  \R(\xbf) := \hat{\R}(\xbf) + \lambda  \overline{\R}(\xbf),
\end{equation}
where $\hat{\R}$ enhances the sparsity of the solution through a learned transform, while $\overline{\R}$ smooths the feature maps non-locally.
Here $\lambda$ is a coefficient to balance these two terms which can be learned. 
The proposed algorithm ELDA to solve \eqref{eq:ls} with regulatization \eqref{eq:r} is computationally efficient, using an inexact gradient approach and only computing safeguard iterates when necessary. Comprehensive convergence and iteration complexity analyses are provided. Numerical experiments show that ELDA, with just 19 layers, outperforms state-of-the-art methods such as RED-CNN \cite{CNN4} and Learned Primal-Dual \cite{adler2018learned}, demonstrating superior reconstruction quality and parameter efficiency.

\subsubsection{Learned Alternating Minimization Algorithm (LAMA) }
Learned Alternating Minimization Algorithm (LAMA) \cite{chi2023} was introduced  for dual-domain sparse-view CT image reconstruction. Induced by a variational model with learnable nonsmooth nonconvex regularizers parameterized as composite functions of deep networks in both image and sinogram domains, LAMA incorporates smoothing techniques and a residual learning architecture. The whole process of the LAMA can be formulated as solving the following minimization problem

\begin{equation}
    \label{eq:OrgPhi}
\argmin_{\mathbf{x},\mathbf{z}}\,\Phi(\mathbf{x},\mathbf{z};\mathbf{s},\Theta)\defeq \frac 1 2 \norm{\*{Ax}-\*z}^2 + \frac \lambda 2 \norm{\*P_{0}\*z - \*s}^2+R(\*x;\theta_1)+Q(\*z;\theta_2).
\end{equation}
Let $(\*x, \*z)$ denote the image and sinogram to be reconstructed, and let $\*s$ represent the sparse-view sinogram. The first two terms in \eqref{eq:OrgPhi}  correspond to the data fidelity and consistency, where $\*A$ denotes the Radon transform and $\*P_0\zbf$ denotes the sparse-view sinogram. Note that $\norm{\cdot} \equiv \norm{\cdot}_2$. The last two terms represent the regularization constraints imposed on the data.

\subsection{Alternating Direction Method of Multipliers (ADMM) inspired optimization network}
PWLS-CSC\cite{bao2019convolutional} addresses the limitations of traditional DL-based computed tomography (CT) reconstruction methods, which are patch-based and often fail to maintain pixel consistency in overlapped patches, leading to redundant features. Convolutional sparse coding (CSC) is utilized to overcome these issues by working directly on the entire image, thus preserving more details and avoiding artifacts from patch aggregation. In PWLS-CSC, a CSC regularization is combined with the penalized weighted least squares (PWLS) model, using filters trained on an external full-sampled CT dataset. Although effective for sparse-view CT, PWLS-CSC may produce ringing artifacts due to inaccurate filters. To mitigate these artifacts, an improved version, PWLS-CSCGR, incorporates gradient regularization on feature maps. Extensive experiments with both simulated and real CT data validate the effectiveness of these methods.

\section{Discussion}
CT reconstruction model in \eqref{eq:ls} can also be applied to MRI reconstruction. For instance, W. Bian et al. \cite{bian2021optimization} developed an MRI reconstruction network within a meta-learning framework, drawing inspiration from an LOA structure. Their experimental results demonstrated promising performance in reconstructing MRI signals from partially sampled k-space data. This work established a foundational network structure for addressing inverse problems and optimization-based models for medical image reconstruction.

\subsection{Future Works}
Recent advances in deep learning have significantly impacted various fields. Deep learning and machine learning techniques are applied across these domains to improve accuracy, efficiency, and capability. For image processing, methods are used for reconstruction and artistic transformations. In LLMs, they enhance understanding and interaction in various contexts, including healthcare and finance. For stock price prediction, advanced time series models and specialized LSTM architectures are employed to capture and predict complex patterns \cite{zhou2024reconstruction,bian2022optimal,yang2024augmentation,bian2024multi,ni2024timeseries,bian2022optimization,ding2024style,bian2023magnetic,10594605,yu2024enhancing,bian2024improving,yu2024advanced,bian2024review,fan2024advanced,bian2022learnable,fan2024towards,wang2024adapting}. CNNs remain crucial for tasks like classification and segmentation, with new architectures such as ResNet and EfficientNet enhancing performance. GANs are used for high-quality image generation and style transfer, with innovations like StyleGAN improving results. Transformers have revolutionized text processing. Models like GPT-4 and T5 offer cutting-edge performance in generating and understanding text. Fine-tuning these models for specific tasks, such as sentiment analysis or medical question answering, has become common.  Long Short-Term Memory (LSTM) networks are used to forecast prices by capturing temporal patterns. Recent improvements include Bidirectional LSTM and attention mechanisms for better accuracy. Transformers also contribute to time series forecasting, modeling complex patterns in stock data. Hybrid models combine deep learning with traditional methods for robust predictions, and reinforcement learning is being explored for optimizing trading strategies and portfolio management.

Deep learning models can be trained on large datasets to recognize patterns, detect objects, segment regions, and perform tasks like denoising or super-resolution. Unlike traditional image processing techniques that rely on manual feature extraction, deep learning approaches learn features directly from the data\cite{li2023deception,bian2024brief,li2019segmentation,zhu2023demonstration}, enabling more accurate and robust image analysis. This has revolutionized fields such as computer vision, medical imaging, and autonomous systems\cite{huang2024ar,bian2020deep,song2024looking,bian2024reviewEMI,song2023going,kang20216,kang2022tie,bian2024optimal,yang2022retargeting}. Deep learning models can be used for real-time object detection, recognizing and tracking the physical props or AR souvenirs as they interact with the environment help process depth information and ensure precise retargeting by understanding visual inputs from the environment and accurately mapping interactions. 

\section{Conclusion}
This paper has explored the recent advancements in learnable optimization-based algorithms (LOA) for low-dose CT reconstruction, addressing the critical need to minimize radiation exposure while maintaining diagnostic image quality. We have discussed several innovative approaches that integrate deep learning within variational models to enhance the regularization process and improve reconstruction quality.
Key developments include:
1. Dual-domain methods like LEARN++ and MAGIC, which leverage complementary information from both image and sinogram domains.
2. Proximal gradient descent-inspired networks such as LDA-LS and ELDA, which offer efficient solutions with theoretical guarantees.
3. ADMM-inspired optimization networks like PWLS-CSC, which address limitations of traditional patch-based methods.

These learnable methods have demonstrated superior performance compared to traditional techniques, offering enhanced artifact reduction, better detail preservation, and robust performance in clinical scenarios. The integration of deep learning within optimization frameworks provides a promising direction for solving ill-posed inverse problems in medical imaging.
Looking ahead, the field of CT reconstruction continues to evolve alongside broader developments in deep learning and artificial intelligence. Future work may focus on:

Improving the interpretability and theoretical foundations of deep learning-based reconstruction methods.
Exploring the potential of emerging AI technologies, such as transformers and large language models, in medical image reconstruction.
Developing more efficient and clinically applicable algorithms that can be seamlessly integrated into existing CT workflows.

As these learnable optimization-based algorithms continue to mature, they hold great promise for revolutionizing low-dose CT imaging, ultimately benefiting patient care by providing high-quality diagnostic images with reduced radiation exposure.
%\subsection* {Acknowledgments}
%This unnumbered section is used to identify those who have aided the authors in understanding or accomplishing the work presented and to acknowledge sources of funding. 

%%%%% References %%%%%
\appendix    % this command starts appendixes

\bibliography{report}   % bibliography data in report.bib
\bibliographystyle{spiejour}   % makes bibtex use spiejour.bst

%%%%% Biographies of authors %%%%%

% \vspace{2ex}\noindent\textbf{Wanyu Bian} is an Image Reconstruction Scientist at Neuro42, specializing in mathematical problems of image and signal processing as well as MRI reconstruction using a blend of optimization and deep learning techniques. In this role, she has pioneered the design of a magnet array optimization algorithm, advancing the field significantly. Previously, she enhanced her expertise as a Research Fellow at Harvard Medical School, focusing on complex challenges in medical imaging. Her academic foundation includes a PhD in Applied Mathematics from the University of Florida, which provided her with deep insights into machine learning, signal processing, MRI, and quantitative imaging.

\end{spacing}
\end{document}